\documentstyle[preprint,aps,graphicx]{revtex}

\begin{document}
\draft
\title{{Phenomenological Analysis of Lepton and Quark Mass Matrices}}

\author{H. NISHIURA}
\address{
Department of General Education, 
Junior College of Osaka Institute of Technology, \\
Asahi-ku, Osaka,535-8585 Japan}
\author{K. MATSUDA, T. KIKUCHI and T. FUKUYAMA}
\address{
Department of Physics, 
Ritsumeikan University, Kusatsu, Shiga, 525-8577 Japan}

\date{February 20, 2002}
\maketitle

\begin{abstract}

We propose a model that all quark and lepton mass matrices originally have the 
same zero texture.
Namely their (1,1), (1,3) and (3,1) components are zeros.
For the neutrino mass matrix, we further impose the symmetry between 
the second and the third generations. 
Then the neutrino mass matrix has a maximal \(\nu_\mu\) - \(\nu_\tau\) 
mixing. 
Our model is consistent with all the neutrino oscillation experiments 
and the experimental data in the quark sector too. 
The neutrino mass matrix can also be well incorporated with 
the seesaw mechanism. 
\end{abstract}
\pacs{PACS number(s): 12.15.Ff, 14.60.Pq, 14.65.-q}

The recent neutrino oscillation experiments\cite{skamioka} have brought us 
much knowledge of the neutrino masses and the lepton mixing. 
In this stage, phenomenological construction of quark and 
lepton mass matrices can be an important step toward the understanding of 
high energy physics beyond the standard model.
In our previous paper\cite{Nishiura3}, we have proposed a mass matrix model 
in which all quark and lepton's mass matrices have the same following texture  
(see also \cite{Ramond}-\cite{Chkareuli2});
\begin{equation}
{\normalsize M=}\left(
	\begin{array}{ccc}
	0 & A & 0\\
	A & B & C\\
	0 & C & D
	\end{array}
\right).
\end{equation}
Namely, 
\(M_u\), \(M_d\), and  \(M_e\) 
(mass matrices of up quarks (\(u,c,t\)), down quarks (\(d,s,b\)), 
and 
charged leptons (\(e,\mu,\tau\)), 
respectively) 
have the same zero texture as follows:
\begin{eqnarray}
&&M_{u}=
\left(
	\begin{array}{ccc}
	 0    & A_{u} & 0 \\
	A_{u} & B_{u} & C_{u}\\
	 0    & C_{u} & D_{u}
	\end{array}
\right), 
\quad
M_{d}=
P_d \left(
	\begin{array}{ccc}
	 0  & A_{d} &  0 \\
	A_{d} & B_{d} & C_{d}\\
	 0  & C_{d} & D_{d}
	\end{array}
\right) P_d^\dagger, \nonumber\\
&&M_{e}=
P_e \left(
	\begin{array}{ccc}
	 0  & A_{e} &  0 \\
	A_{e} & B_{e} & C_{e}\\
	 0  & C_{e} & D_{e}
	\end{array}
\right) P_e^\dagger \label{eq5}.
\end{eqnarray}
where \(P_d\equiv\mbox{diag}(e^{i\alpha_1},e^{i\alpha_2},e^{i\alpha_3})\), 
\(\alpha_{ij}\equiv \alpha_i-\alpha_j\), and 
 \(P_e\equiv\mbox{diag}(e^{i\beta_1},e^{i\beta_2},e^{i\beta_3})\), 
\(\beta_{ij}\equiv \beta_i-\beta_j\) are the $CP$ violating phase factors.
In Ref\cite{Nishiura3}, we also assumed that 
the mass matrix \(M_\nu\) of neutrinos (\(\nu_e,\nu_\mu,\nu_\tau\)) has 
the same texture, 
\begin{eqnarray}
\nonumber \\
&&M_{\nu}=
\left(
	\begin{array}{ccc}
	 0  & A_{\nu} &  0 \\
	A_{\nu} & B_{\nu} & C_{\nu}\\
	 0  & C_{\nu} & D_{\nu}
	\end{array}
\right).
\label{eq10}
\end{eqnarray}
In this paper, however, being motivated by the maximal atmospheric 
neutrino mixing\cite{skamioka}, we further impose the symmetry for 
\(M_\nu\) under the permutation of the second and the third generations 
in place of the form of Eq.(\ref{eq10}).
That is, first let us decompose the \(M_\nu\) of Eq.(\ref{eq10}) into 
the symmetric and anti-symmetric parts 
with respect to the permutation of the second and the third generations,
\begin{equation}
\left(
	\begin{array}{ccc}
	0 & A_{\nu} & 0\\
	A_{\nu} & B_{\nu} & C_{\nu}\\
	0 & C_{\nu} & D_{\nu}
	\end{array}
\right)=
\left(
	\begin{array}{ccc}
	0 & \frac{1}{2}A_{\nu}& \frac{1}{2}A_{\nu}\\
	\frac{1}{2}A_{\nu} & \frac{1}{2}(D_{\nu}+B_{\nu}) & C_{\nu}\\
	\frac{1}{2}A_{\nu} & C_{\nu} & \frac{1}{2}(D_{\nu}+B_{\nu})
	\end{array}
\right)+
\left(
	\begin{array}{ccc}
	0 & \frac{1}{2}A_{\nu} & -\frac{1}{2}A_{\nu}\\
	\frac{1}{2}A_{\nu} & -\frac{1}{2}(D_{\nu}-B_{\nu}) & 0\\
	-\frac{1}{2}A_{\nu} & 0 & \frac{1}{2}(D_{\nu}-B_{\nu})
	\end{array}
\right).
\end{equation}
Then, by imposing the above symmetry on \(M_\nu\), 
we keep only the symmetric part,
\begin{equation}
M_{\nu}=
\left(
	\begin{array}{ccc}
	0 & A_{\nu} & 0\\
	A_{\nu} & B_{\nu} & C_{\nu}\\
	0 & C_{\nu} & D_{\nu}
	\end{array}
\right)_{2\leftrightarrow3 sym. part}
=
\left(
	\begin{array}{ccc}
	0 & \frac{1}{2}A_{\nu}& \frac{1}{2}A_{\nu}\\
	\frac{1}{2}A_{\nu} & \frac{1}{2}(D_{\nu}+B_{\nu}) & C_{\nu}\\
	\frac{1}{2}A_{\nu} & C_{\nu} & \frac{1}{2}(D_{\nu}+B_{\nu})
	\end{array}
\right)
\equiv
\left(
	\begin{array}{ccc}
	0 & Y_{\nu} & Y_{\nu}\\
	Y_{\nu} & Z_{\nu} & W_{\nu}\\
	Y_{\nu} & W_{\nu} & Z_{\nu}
	\end{array}
\right). \label{eq15}
\end{equation}
The structure of Eq.(\ref{eq15}) was previously suggested in Ref\cite{Fukuyama}-\cite{Grimus} 
using the basis where the charged-lepton mass matrix is diagonal.
For \(M_u\), \(M_d\), and  \(M_e\), 
we keep both the symmetric and anti-symmetric parts of them. 
The Eqs.(\ref{eq5}) and (\ref{eq15}) are our new proposed mass matrix forms for quarks and leptons.
\par
In this paper, in order to get the neutrino mass eigenvalues, we assign 
\begin{eqnarray}
Y_{\nu}& = \pm &\sqrt{\frac{m_1m_2}{2}},\quad 
Z_{\nu} = \frac{1}{2}(m_3+m_2-m_1),\nonumber\\
W_{\nu}& = &-\frac{1}{2}(m_3-m_2+m_1),
\end{eqnarray}
where \(m_1\), \(m_2\) and \(m_3\) are the neutrino masses.
Then the \(M_{\nu}\) is diagonalized by an orthogonal matrix \(O_{\nu}\) as
\begin{equation}
O_{\nu}^T
\left(
	\begin{array}{ccc}
	0 & Y_{\nu} & Y_{\nu}\\
	Y_{\nu} & Z_{\nu} & W_{\nu}\\
	Y_{\nu} & W_{\nu} & Z_{\nu}
	\end{array}
\right)
O_{\nu}=
\left(
	\begin{array}{ccc}
	-m_1 & 0 & 0\\
	0 & m_2 & 0\\
	0 &  0 & m_3
	\end{array}
\right),
\end{equation}
with 
\begin{equation}
O_{\nu} = 
\left(
	\begin{array}{ccc}
 	\mp \scriptstyle{\sqrt{\frac{m_2}{(m_2+m_1)}}}& \qquad 
 	\pm \scriptstyle{\sqrt{\frac{m_1}{(m_2+m_1)}}}&  \qquad
	0\\
	\scriptstyle{\sqrt{\frac{m_1}{2(m_2+m_1)}}}&  \qquad
	\scriptstyle{\sqrt{\frac{m_2}{2(m_2+m_1)}}}&  \qquad
	\scriptstyle{-\frac{1}{\sqrt{2}}}\\
	\scriptstyle{\sqrt{\frac{m_1}{2(m_2+m_1)}}}&  \qquad
	\scriptstyle{\sqrt{\frac{m_2}{2(m_2+m_1)}}}&  \qquad
	\scriptstyle{\frac{1}{\sqrt{2}}}
	\end{array}
\right).
\end{equation}
Note that the elements of \(O_{\nu}\) are independent of \(m_3\) 
because of the above structure of \(M_\nu\). 
\par 
Next let us discuss the mass matrix \(M_e\) of the charged leptons.
Using the type I set discussed in Ref\cite{Nishiura3}, 
we assign  
\begin{eqnarray}
A_e& =&\sqrt{\frac{m_e m_\mu m_\tau}{m_\tau-m_e}}, \qquad
B_e=m_\mu,\nonumber\\
C_e& =&\sqrt{\frac{m_e m_\tau(m_\tau-m_\mu-m_e)}{m_\tau-m_e}}, \qquad
D_e=m_\tau-m_e. \label{eq19}
\end{eqnarray}
Here \(m_e\), \(m_\mu\) and \(m_\tau\) are the charged lepton masses.
Then mass matrix \(M_e\) becomes
\begin{eqnarray}
&&M_e \simeq
P_e\left(
	\begin{array}{ccc}
	0& \sqrt{m_em_\mu}& 0\\
	\sqrt{m_em_\mu} & m_\mu & \sqrt{m_em_\tau}\\
	0& \sqrt{m_em_\tau} & m_\tau-m_e
	\end{array}
\right)P_e^\dagger \label{eq20}\\
&&\hspace{5cm} (\mbox{for }m_\tau \gg m_\mu \gg m_e)\nonumber.
\end{eqnarray}
The \(M_e\) is diagonalized by \(P_e O_e\). 
Here the orthogonal matrix \(O_e\) which diagonalizes
\(P_e^\dagger M_e P_e\) is obtained as
\begin{equation}
O_e^T
\left(
	\begin{array}{ccc}
	0 & A_e & 0\\
	A_e & B_e & C_e\\
	0 & C_e & D_e
	\end{array}
\right)
O_e=
\left(
	\begin{array}{ccc}
	-m_e & 0 & 0\\
	0 & m_\mu & 0\\
	0 &  0 & m_\tau
	\end{array}
\right),
\end{equation}
with \(O_e\) given, for \(m_\tau \gg m_\mu \gg m_e\), by
\begin{equation}
O_e \simeq
\left(
	\begin{array}{ccc}
 	1& \sqrt{\frac{m_e}{m_\mu}}&
	 \sqrt{\frac{m_e^2m_\mu}{m_\tau^3}}\\
	-\sqrt{\frac{m_e}{m_\mu}}
	& 1 & \sqrt{\frac{m_e}{m_\tau}}\\
	\sqrt{\frac{{m_e^2}}{m_\mu m_\tau}}
	& -\sqrt{\frac{m_e}{m_\tau}} & 1
	\end{array}
\right)\label{eq30}.
\end{equation}
In this case, the Maki-Nakagawa-Sakata (MNS)\cite{MNS} lepton mixing matrix 
\(U\) is given by
\begin{equation}
U=P_l^\dagger P_e^\dagger O_e^T P_e O_\nu P_l=
 \left(
 	\begin{array}{ccc}
 	U_{e1} & U_{e2} & U_{e3}\\
 	U_{\mu 1} & U_{\mu 2} & U_{\mu 3}\\
 	U_{\tau 1} & U_{\tau 2} & U_{\tau 3}
 	\end{array}
 \right),
\end{equation}
where \(P_l=\mbox{diag}(i,1,1)\) is included to have positive neutrino masses.
The \(P_l^\dagger P_e^\dagger\) factor leads \(U\) to the form whose diagonal
elements are real to a good approximation. 
We obtain the expressions of some elements of \(U\) 
by keeping terms up to order of \(\sqrt{\frac{m_e}{m_\mu}}\) as follows:
\begin{eqnarray}
&&U_{e1}\simeq
\mp\sqrt{\frac{m_2}{(m_2+m_1)}}
-\sqrt{\frac{m_1}{(m_2+m_1)}}\sqrt{\frac{m_e}{m_\mu}}e^{-i \beta_{12}},
\nonumber\\
&&U_{e2}\simeq
-i\left(
\pm\sqrt{\frac{m_1}{(m_2+m_1)}}
-\sqrt{\frac{m_2}{(m_2+m_1)}}\sqrt{\frac{m_e}{m_\mu}}e^{-i \beta_{12}}
\right),
\nonumber\\
&&U_{e3}\simeq
-i
\frac{1}{\sqrt{2}}\sqrt{\frac{m_e}{m_\mu}}e^{-i \beta_{12}}.\label{eq40}
\end{eqnarray}
Here we neglect the terms of order of \(\sqrt{\frac{m_e^2}{m_\mu m_\tau}}\).
In the leading order with respective to 
\(\frac{m_e}{m_\mu}\), we have
\begin{equation}
U \simeq 
\left(
	\begin{array}{ccc}
 	\mp \scriptstyle{\sqrt{\frac{m_2}{(m_2+m_1)}}}& 
 	\pm (-i)\scriptstyle{\sqrt{\frac{m_1}{(m_2+m_1)}}}&  
	-i\frac{1}{\sqrt{2}}\sqrt{\frac{m_e}{m_\mu}}e^{-i \beta_{12}}\\
	i\scriptstyle{\sqrt{\frac{m_1}{2(m_2+m_1)}}}&  
	\scriptstyle{\sqrt{\frac{m_2}{2(m_2+m_1)}}}& 
	\scriptstyle{-\frac{1}{\sqrt{2}}}\\
	i\scriptstyle{\sqrt{\frac{m_1}{2(m_2+m_1)}}}& 
	\scriptstyle{\sqrt{\frac{m_2}{2(m_2+m_1)}}}&  
	\scriptstyle{\frac{1}{\sqrt{2}}}
	\end{array}
\right).\label{eq50}
\end{equation}
From Eq.(\ref{eq50}), we obtain a maximal atmospheric neutrino mixing angle
\begin{equation}
\theta_{23} \simeq\frac{\pi}{4}.\label{eq55}
\end{equation}
The averaged neutrino mass \(\langle m_{\nu} \rangle\equiv |\sum _{j=1}^{3}U_{ej}^2m_j| \) 
defined from the neutrinoless 
double beta decay\cite{doi} is also obtained from Eq.(\ref{eq40}), as
\begin{equation}
\langle m_{\nu} \rangle = \left|(\frac{m_3}{2}+m_2-m_1)\frac{m_e}{m_\mu}
\mp2\sqrt{m_1m_2}\sqrt{\frac{m_e}{m_\mu}}e^{i \beta_{12}}\right|.
\label{betabetamass}
\end{equation}
It should be noted that \(m_1\), \(m_2\), and \(m_3\) remain free parameters 
and can be fixed from the solar and atmospheric neutrino data, $|U_{e2}|^2$, $\Delta m_{\mbox{{\tiny solar}}}^2$, and $\Delta m_{\mbox{{\tiny atm}}}^2$ 
as well as the averaged neutrino mass from neutrinoless double beta decay, 
\(\langle m_{\nu} \rangle\).
From Eqs.(\ref{eq40}) and (\ref{betabetamass}), with varying the unknown $CP$ violating phase \(\beta_{12}\), we obtain the following constraints on \(|U_{e2}|^2\) and \(\langle m_{\nu} \rangle\):
\begin{eqnarray}
\left|\sqrt{\frac{m_1}{m_2+m_1}}
-\sqrt{\frac{m_2}{m_2+m_1}}\sqrt{\frac{m_e}{m_\mu}}\right|^2
\le
|U_{e2}|^2
\le
\left|\sqrt{\frac{m_1}{m_2+m_1}}
+\sqrt{\frac{m_2}{m_2+m_1}}\sqrt{\frac{m_e}{m_\mu}}\right|^2,\label{eq68} \\
\left|(\frac{m_3}{2}+m_2-m_1)\frac{m_e}{m_\mu}-2\sqrt{m_1m_2}\sqrt{\frac{m_e}{m_\mu}}\right|
\le
\langle m_{\nu} \rangle
\le
\left|(\frac{m_3}{2}+m_2-m_1)\frac{m_e}{m_\mu}+2\sqrt{m_1m_2}\sqrt{\frac{m_e}{m_\mu}}\right|. 
\label{eq69}
\end{eqnarray}
\par
In the following discussions we consider the normal mass hierarchy 
\(\Delta m_{23}^2>0\) for the neutrino mass. 
The case of the inverse mass hierarchy 
\(\Delta m_{23}^2<0\) is quite similar to that of the normal mass hierarchy. Substituting \(m_e=0.51\)MeV and \(m_\mu=106\)MeV into Eq.(\ref{eq40}), we predict
\begin{equation}
|U_{e3}|^2 = 0.0024,
\end{equation}
which is consistent with the experimental constraint 
$|U_{e3}|_{\mbox{\tiny exp.}}^2 <  0.03$ obtained from the CHOOZ\cite{chooz}, solar, and atmospheric neutrino experiments\cite{skamioka}.
Let us estimate the neutrino mass \(m_i\) 
by fitting the experimental data.
From the solar neutrino experiment, we have
\begin{eqnarray} 
\Delta m_{12}^2=m_2^2-m_1^2=\Delta m_{\mbox{{\tiny solar(MSW)}}}^2=1.0\times 10^{-5}\mbox{eV}^2, \\
0.3\le|U_{e2}|^2\le0.7   \quad\mbox{for LMA-MSW}, \label{eq20501} \\
1\times 10^{-3}\le|U_{e2}|^2\le1\times 10^{-2}   \quad\mbox{for SMA-MSW}, \label{eq20502}
\end{eqnarray} 
for the large mixing angle(LMA) and the small mixing angle(SMA) MSW solutions.
From the atmospheric neutrino experiment\cite{skamioka}, we also have
\begin{equation}
\Delta m_{23}^2=m_3^2-m_2^2=\Delta m_{\mbox{{\tiny atm}}}^2=(1-7)\times 10^{-3}\mbox{eV}^2. 
\end{equation} 
Combining these experimental data with Eq.(\ref{eq68}), 
we obtain the allowed regions in the \(m_2^2\)-\(m_1^2\) plane, which are shown in Fig. 1. 
The allowed regions in the $|U_{e2}|^2$-$m_1$ plane are shown in Fig. 2.
From Eq.(\ref{eq69}), we also obtain the allowed regions in the $\langle m_{\nu} \rangle$-$m_1$ plane as shown in Fig. 3.
It turns out from Fig. 2 and Fig. 3 that large $\langle m_{\nu} \rangle$ prefers the large mixing angle MSW solution for the solar neutrino problem.
From Fig. 3 and the recent experimental 
upper bound $\langle m_{\nu} \rangle<0.2$ eV\cite{baudis}, 
we obtain the upper bound for \(m_1\) as 
\(m_1 \alt 1.5 \mbox{ eV}\).
If we impose the constraint of neutrino masses from 
the astrophysical observation \cite{fukugita}, $\sum_i m_i < 1.8$ eV, 
on Fig. 3, we obtain 
\begin{equation}
\langle m_{\nu} \rangle \alt 0.08 \mbox{ eV}.\label{eq85}
\end{equation}
Recently, Klapdor-Kleingrothaus et al.\cite{Klapdor} have discussed 
the evidence for neutrinoless double beta decay. 
If this is true, it is a big news. 
They have reported  \(\langle m_\nu \rangle\)\(=\)\((0.05-0.84)\)eV (95$\%$ C.L.),
which is consistent with Eq.(\ref{eq85}).
\par
The mass matrices for quarks \(M_d\) and \(M_u\) in Eq.(\ref{eq5}) are 
the same as our previous model and the phenomenological results of them have been discussed in Ref\cite{Nishiura3}. 
So we discuss the mass matrices for quarks only briefly. 
The \(M_d\) and \(M_u\) are assumed to be of type I, 
similarly to the charged leptons given in Eqs.(\ref{eq19}) and (\ref{eq20}), as 
\begin{eqnarray}
M_d &\simeq&
P_d
\left(
	\begin{array}{ccc}
	0              &  \sqrt{m_d m_s}  &  0\\
	\sqrt{m_d m_s} &  m_s               &  \sqrt{m_d m_b}\\
	0              &  \sqrt{m_d m_b}  &  m_b-m_d
	\end{array}
\right)P_d^\dagger, \nonumber \\
M_u &\simeq&
\left(
	\begin{array}{ccc}
	0              &  \sqrt{m_u m_c}  &  0\\
	\sqrt{m_u m_c} &  m_c               &  \sqrt{m_u m_t}\\
	0              &  \sqrt{m_u m_t}  &  m_t-m_u
	\end{array}
\right),
\end{eqnarray}
where \(m_d\), \(m_s\), and \(m_b\) are down quark masses and 
\(m_u\), \(m_c\), and \(m_t\) are up quark masses.
The Cabbibo-Kobayashi-Maskawa (CKM) \cite{CKM} quark mixing matrix 
derived from those \(M_d\) and \(M_u\) is consistent with 
the experimental data.(See Ref\cite{Nishiura3} for details.)
\par
Finally, let us incorporate the seesaw mechanism in our model.
In the seesaw mechanism\cite{Yanagida}, the neutrino mass matrix 
\(M_\nu\) is given by 
\begin{equation}
M_\nu=-M_D^T M_R^{-1} M_D.\label{eq70} 
\end{equation}
Here \(M_D\) is the Dirac neutrino mass matrix and \(M_R\) is the Majorana
mass matrix of the right-handed components. 
In this paper, we furthermore assume that the \(M_D\) and \(M_R\) have the same zero texture as \(M_\nu\) of Eq.(\ref{eq15}):  
\begin{eqnarray}
M_D
& = &
\left(
	\begin{array}{ccc}
	0 & A_D & 0\\
	A_D & B_D & C_D\\
	0 & C_D & D_D
	\end{array}
\right)_{2\leftrightarrow3 sym. part}
=
\left(
	\begin{array}{ccc}
	0 & Y_D & Y_D\\
	Y_D & Z_D & W_D\\
	Y_D & W_D & Z_D
	\end{array}
\right),\nonumber\\
M_R
& = &
\left(
	\begin{array}{ccc}
	0 & A_R & 0\\
	A_R & B_R & C_R\\
	0 & C_R & D_R
	\end{array}
\right)_{2\leftrightarrow3 sym. part}
=
\left(
	\begin{array}{ccc}
	0 & Y_R & Y_R\\
	Y_R & Z_R & W_R\\
	Y_R & W_R & Z_R
	\end{array}
\right).
\end{eqnarray}
This form conserves its form via the seesaw mechanism as discussed 
in Ref\cite{Fukuyama}. 
Namely, the \(M_\nu\) given by Eq.(\ref{eq70}) takes the same form too 
and is given by Eq.(\ref{eq15}). 
\par
Conclusive remarks are in order.
We have started with the same type of 4 texture zero mass matrices 
both for quarks and leptons. 
For the neutrino mass matrix we have further imposed the symmetry between 
the second and the third generations. Our new proposed mass matrices 
for quarks and leptons are given by 
Eqs.(\ref{eq5}) and (\ref{eq15}) 
which are consistent with present experiments. We have also shown that 
the neutrino mass matrix can be well incorporated with 
the seesaw mechanism. 
\ \\

The work of K.M. is supported by the JSPS Research Grant No. 10421.



\begin{figure}[htbp]
\begin{center}
\includegraphics{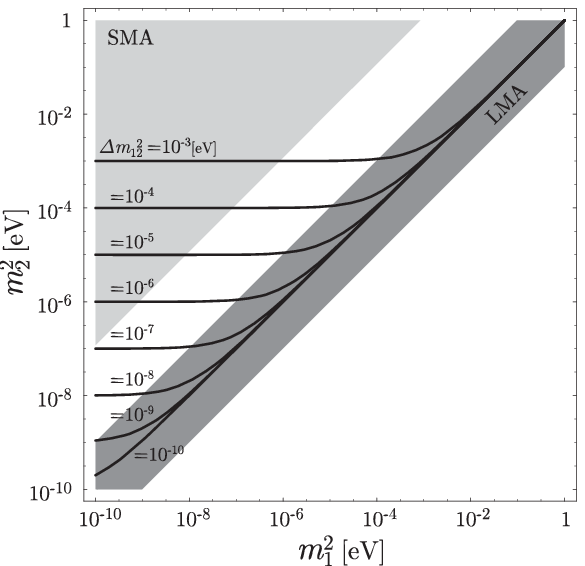}
\end{center}
\caption{The allowed regions of \(m_2^2\) and \(m_1^2\). 
The dark- and light- shaded regions are allowed for LMA and SMA solutions, Eqs.(\ref{eq20501}) and (\ref{eq20502}), respectively. The curve passing inside of those shaded regions is allowed for each fixed \(\Delta m_{12}^2\).}
\label{fig1}
\end{figure}
\begin{figure}[htbp]
\begin{center}
\includegraphics{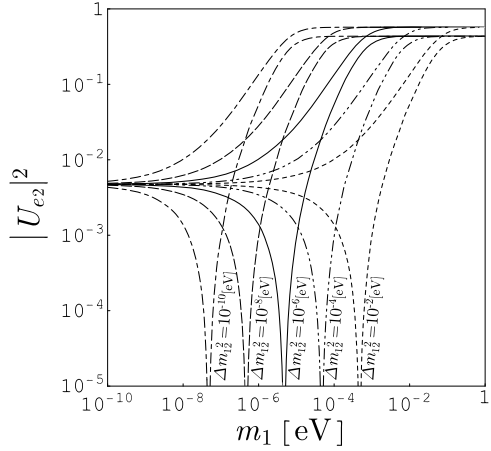}
\end{center}
\caption{The allowed regions of \(|U_{e2}|^2\) and \(m_1\). 
The allowed regions are inside of the boundary curves for each fixed \(\Delta m_{12}^2\). }
\label{fig2}
\end{figure}


\begin{figure}[htbp]
\begin{center}
\includegraphics{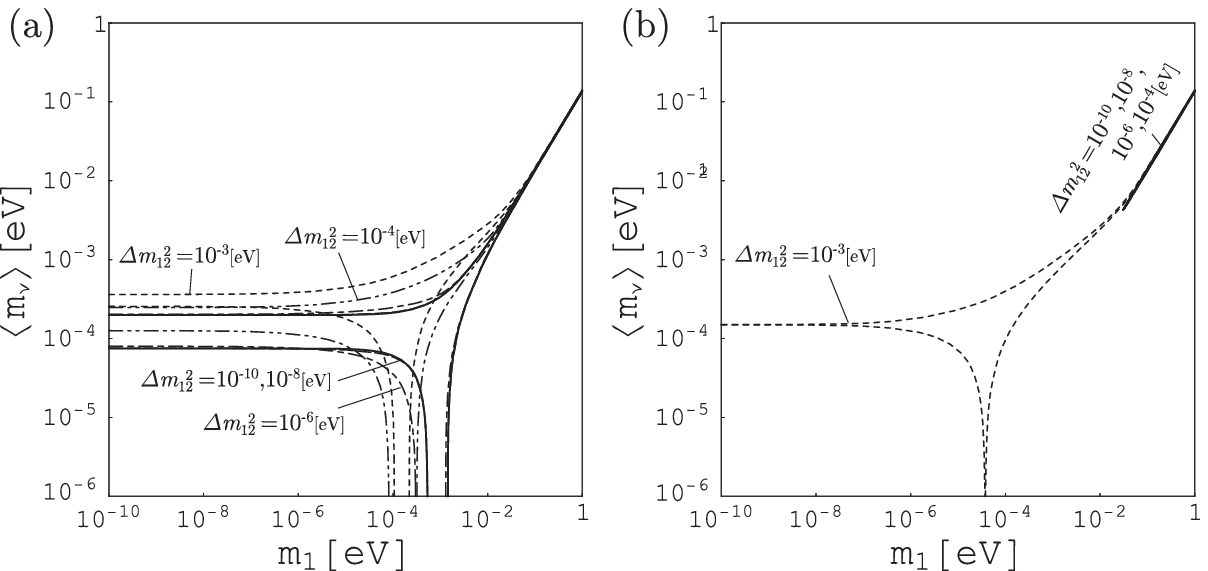}
\end{center}
\caption{The allowed regions of \(\langle m_\nu \rangle\) and \(m_1\). 
The allowed regions are inside of the boundary curves for each fixed \(\Delta m_{12}^2\).
(a) Normal neutrino mass hierarchy case \(\Delta m_{23}^2=+(1-7)\times10^{-3}\) eV\(^2\).
(b) Inverse neutrino mass hierarchy case \(\Delta m_{23}^2=-(1-7)\times10^{-3}\) eV\(^2\).
 }
\label{fig3}
\end{figure}

\end{document}